\documentstyle[aps,multicol,prl,epsf]{revtex}
\def\d{{\rm d}}
\begin{document}
\title{Collapse of Stiff Polyelectrolytes due to Counterion 
Fluctuations}
\author{Ramin Golestanian$^{1,2,3,5}$, 
Mehran Kardar$^{2}$, and Tanniemola B. Liverpool$^{3,4,5}$}
\address{$^{1}$Institute for Advanced Studies in Basic Sciences,
Zanjan 45195-159, Iran
\\ $^{2}$Department of Physics, Massachusetts Institute of Technology, 
Cambridge, 
MA 02139
\\ $^{3}$Max-Planck-Institut f{\"u}r Polymerforschung,
D-55021 Mainz, Germany
\\ $^{4}$Physico-Chimie Theorique, ESA CNRS 7083, E.S.P.C.I., 75231 Paris Cedex 
05, France
$^{*}$
\\ $^{5}$ Institute for Theoretical Physics, University of California,
Santa Barbara, CA 93106-4030}
\date{\today}
\maketitle
\begin{abstract}

The effective elasticity of highly charged stiff polyelectrolytes is studied in  
the presence of counterions, with and without added salt.
The rigid polymer conformations may become unstable due to an effective 
attraction induced by counterion density fluctuations. 
Instabilities at the longest, or intermediate length scales may signal
collapse to globule, or necklace states, respectively.
In the presence of added-salt, a generalized electrostatic persistence 
length is obtained, which has a nontrivial dependence on
the Debye screening length.
\end{abstract}
\begin{multicols}{2}
A polyelectrolyte (PE) is an ionic polymer which when dissolved in polar 
solvents dissociates into a long polymer chain (macroion), and small mobile 
{\it counterions}. 
Because of the electrostatic repulsion of the uncompensated charges
on the polymer, the chain is stretched out to rod-like conformations.
Naively, one might expect that the more highly charged PEs are the more stiff.
However, this tendency is opposed by the stronger attraction to the
counterions, which may condense on a highly charged PE, giving it a 
much lower apparent charge\cite{Man,Oos}, and resulting in a lower stiffness. 
Nonetheless, in a mean-field (Poisson-Boltzmann) treatment of the
counterions, the conformation of the PE is stretched.

On the other hand, it is known experimentally that both highly charged 
flexible PEs (such as polystyrene sulphonate) \cite{Delsanti,Monica}, 
and stiff PEs (such as DNA)\cite{Bloomfield}, can collapse 
in the presence of multivalent counterions to highly compact states. 
Rodlike PEs also  form bundles under similar 
conditions\cite{Bundle}. To account for these effects, a relatively 
long-ranged attractive interaction capable of competing with
the residual coulomb repulsion (of the PE backbones with 
compensated charge densities) is needed. 

To understand the origin of the attractive interaction, consider
two like-charged substrates on which counterions are condensed. 
As the two objects approach each other, the counterions may
rearrange their positions.
Any ``correlated separation of charges'' now leads to an {\it attraction}
whose range is of the order of the ``correlation hole'' of
the counterions on each substrate. 
Several different mechanisms may lead to correlated charge separation: 
At low temperatures, the counterions form
a Wigner crystal on the charged substrate, which leads to an attractive interaction
with a range set by the lattice spacing \cite{Rouzina2,Levin}. 
Another mechanism for charge separation is a specific binding of 
counterions to the substrate\cite{Leikin}.
In this case the periodicity, and hence the range of the attraction, 
is dictated by the underlying structure of the substrate. 
Thermal fluctuations can also induce instantaneous charge separations 
which inter-correlate on the two objects, 
leading to an attraction similar to the van der Waals interaction. 
Since dominant contributions come from fluctuations with wavelengths
of the order of the separation of the objects, the range of the attraction
is set by the distance between them; i.e. the interaction is long-ranged
\cite{Oos,RodFFI,HaLiu,KGRMP}.

Recently, there have been a number of studies of the role of counterion 
condensation in the transition between extended and collapsed states of 
{\it flexible} PEs \cite{Monica,SK,BKK}. 
The situation is more subtle for {\it stiff} PEs due to their intrinsic rigidity 
\cite{Man2,Rouzina,Podgornik}. 
Here we employ  path integral methods\cite{KGRMP,LiK,long} to 
study the energy cost of deforming a stiff and highly charged PE in 
the presence of thermally fluctuating counterions. 
In particular, consider a chain of total charge $N_c$ and length $L$,
with a microscopic persistence length $\ell_p$, 
and average separation  $a$ between charges on its backbone, 
in a neutralizing solution of counterions of valence $z$. 
The PE is highly charged when $a$ is less than the Bjerrum length 
$\ell_B=e^2/\epsilon k_B T$, where $\epsilon$ is the dielectric constant of the solvent. 
(For water at room temperature, $\ell_B\simeq 7.1 \AA$.)
We calculate the effective free energy of a fluctuating PE as a 
perturbative expansion in the deformations around an average rodlike structure. 
The linear stability of this structure is controlled by a spectrum ${\cal E}(k)$, 
as a function of the deformation wavevector $k$.
A negative value of ${\cal E}(k)$ signals an instability at the corresponding
wavelength, leading to phase diagrams as in Figs.~2 and 3.
In particular, in the absence of salt (Fig.~2), we find that counterion fluctuations 
{\it cannot} trigger collapse of a stiff  PE, {\it unless} it has a microscopic 
persistence length less than a critical value of
\begin{equation}
\ell_p^c=\Delta \times a \times z^4 \times \left({\ell_B \over a}\right)^3
                \times \left(1-{a \over z \ell_B}\right)^2,        
       \label{lc1}
\end{equation}  
in which $\Delta$ is a numerical constant given later. 
For $\ell_p < \ell_p^c$, there is a finite domain of intermediate lengths $L$
of the PE for which a collapsed conformation is favored.

For a low concentration of added salt, we find an effective 
persistence length of 
\begin{equation}
L_p=\ell_p+\frac{a \left(a / \ell_B \right)^3}{16 z^4 
 \left(1-{a /  z \ell_B}\right)^2 (\kappa a)^2 
\ln^2\left({1 \over \kappa a}\right)}
-\frac{c_2}{2 \kappa \ln\left(1 \over \kappa a\right)},  \label{Lpeff}
\end{equation}
where $\kappa^{-1}$ is the Debye screening length, related to
the salt (number) density $n$ via $\kappa^2=4 \pi \ell_B n$, 
and $c_2$ is a numerical constant given below. 
The above expression, which is a {\it non-trivial} generalization of the
Odijk-Skolnick-Fixman (OSF) electrostatic persistence length \cite{OSF}, 
is a sum of repulsive (+) and attractive (-) electrostatic contributions. 
It can therefore be negative under certain conditions, which we take 
as an indication of a conformational instability of the 
rodlike PE (tendency to collapse). 
This occurs in the regions of the phase diagram indicated in  Fig.~3,
and only for persistence lengths less than a critical value given by an
expression similar to Eq.(\ref{lc1}). 
A similar `softening' contribution to the rigidity of charged membranes, 
and associated conformational instability, is reported in Ref.\cite{LP}.
There are related instabilities caused by surface fluctuation-induced 
interactions between stiff polymers on membranes\cite{RG}.

\begin{figure}
\centerline{
\epsfxsize=7truecm
\epsffile{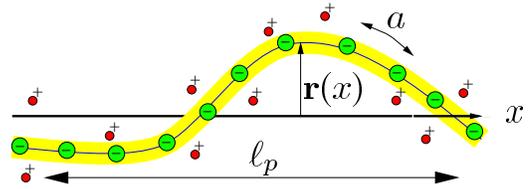}
}
\vskip 0.3truecm
\caption{Transverse deformations of a rodlike PE.\hskip12truecm
}
\label{conf}
\end{figure}

Without loss of generality, we orient the PE along the $x$-axis, 
and parametrize its configuration by a 3D vector $ R(x)$. 
For small deformations (no overhangs) as in Fig.~1, we can set 
$ R(x)=(x,{\bf r}(x))$, where the 2D vector ${\bf r}(x)$ measures 
transverse deviations from a rod for $0\leq x\leq L$.
For each configuration we would like to calculate a constrained partition 
function ${\cal Z}[R]$, by integrating over all counterions positions in solution.
As this is quite difficult, we instead generalize the approach of Ref.\cite{HaLiu}, 
and integrate over a fluctuating counterion charge 
density along the PE backbone, as \cite{long}
\end{multicols}
\begin{equation}
{\cal Z}[R]\equiv e^{- \beta {\cal H}^{eff}_0[R]}=e^{- \beta {\cal H}^p[R]} 
\int {\cal D} q(x) \;\exp\left\{
- {1 \over 2} \int {\d x \over a} \frac{(q(x)-q_0)^2}{(\delta q)^2}  
-{ \ell_B \over 2} \int {\d x \over a}{\d x' \over a} \frac{q(x) q(x')   
e^{-\kappa |R(x)-R(x')|}}{|R(x)-R(x')|}\right\}.       
\label{Z2}
\end{equation}  
\begin{multicols}{2}
\noindent
The first term, $ \beta {\cal H}^p[R]=\ell_p  \int \d x (\partial_x^2 R)^2/2$,
is the energy cost of bending the PE, and  $\beta=1/k_B T$.
In the second term, the condensate charge is integrated over a gaussian
distribution of mean $q_0=a/z \ell_B$, and variance 
$(\delta q)^2=z (1-a/z \ell_B)$\cite{HaLiu,long}. 
A finite $\kappa$ anticipates the effect of added salt\cite{fisher}.

Using the methods of Refs.\cite{KGRMP,LiK}, the effective Hamiltonian 
can be calculated perturbatively in the deformation field ${\bf r}(x)$. 
In terms of the Fourier modes ${\bf r}(k)$, the effective free energy
cost of deformations is
\begin{equation}
\beta {\cal H}^{eff}_0[{\bf r}(x)]={1 \over 2} \int {\d k \over 2 \pi}\;{\cal 
E}(k)\;
|{\bf r}(k)|^2 +O(r^4), \label{Fns} 
\end{equation}
where in the absence of salt  $(\kappa=0)$
\begin{equation}
{\cal E}(k)=
\frac{(q_0^2/a^2) \ell_B\; k^2 \ln(k/k_0)}
{\left[1+2(\delta q)^2(\ell_B/a)\ln(L/a)\right]^2} -B_1 k^3+\ell_p k^4.    \label{E1}
\end{equation}  
(Despite its appearance, Eq.(\ref{E1}) {\it is not} a Taylor expansion in $k$.)
The rigidity term, $\ell_p k^4$, ensures stability of the PE at large $k$ 
(small length-scales). 
The first term describes the stiffening of the PE due to Coulomb
interactions, which is most apparent at long wavelengths, and
also involves a cutoff $k_0 \ll \pi/L$ \cite{k0}.
Note that the fluctuations in charge reduce the strength of this term
considerably, compared to the uniform case with $(\delta q)^2=0$\cite{KK,LiWi}).
The new element is the term $-B_1k^3$ which is due to the
fluctuation-induced attractions\cite{k^3}, and may lead to
instabilities at intermediate $k$.
The coefficient $B_1$ can be obtained as an integral that depends
weakly on $(\delta q)^2(\ell_B/a)$, and that for typical values is well 
approximated as $B_1 \simeq c_1/\ln(L/a)^2$ with $c_1\simeq 0.101$. 

The rodlike configuration is unstable if any ${\cal E}(k)$ is negative.
The onset of instability is thus determined from ${\cal E}(k_m)=0$,
at the minimum of the spectrum given by ${\d{\cal E}(k_m) / \d k}=0$. 
In the limit of $L\gg a$, these conditions can be expressed as
$\ell_p=\ell_p^c$, with the critical persistence length given by 
Eq.(\ref{lc1}) above,  with 
$\Delta_{sf}\approx c_1^2/\ln^2(L/a) \ln[c_1/2 \ell_p k_0 \ln^2(L/a)]$. 
For $\ell_p < \ell_p^c$, there is domain of unstable modes for 
$k_{-} < k < k_{+}$, where $k_{\pm}=
\left[c_1/2\ell_p\ln^2(L/a)\right]\left(1\pm \sqrt{1-\ell_p/\ell_p^c}\,\right)$.
For finite chains, the values of $k$ are restricted to $k>\pi/L$,
leading to the following three possibilities: 
{\bf (i)} For $k_{+} < \pi/L$, the unstable modes cannot be accessed 
and the PE retains its {\it  extended} form. 
{\bf (ii)} For $k_{-} < \pi/L < k_{+}$, the PE is unstable at the longest
scales. It will fold in a manner which we interpret as a precursor 
to collapse into a {\it globular} state. 
{\bf (iii)} For $\pi/L < k_{-}$,  the PE is stable at the longest scales, but
unstable at intermediate lengths. 
Although this analysis does not determine the fate of the unstable PE, 
we guess that this instability signals the formation of a 
{\it necklace} structure\cite{necklace}.
These tentative identifications are depicted in the phase diagram of Fig.~2.
The  boundary of the collapsed phase is obtained from $k_{\pm}=\pi/L$, as
\begin{equation}
{\ell_p \over a}\approx\frac{L/a}{\pi \ln^2(L/a)}\left[c_1
-\frac{(L/a) \ln(\pi/L k_0)}{4\pi z^4 (\ell_B/a)^3 (1-a/z \ell_B)^2}\right],
\label{ph1}
\end{equation}  
and has a nearly parabolic shape.
\begin{figure}
\centerline{\epsfxsize 8cm{\epsffile{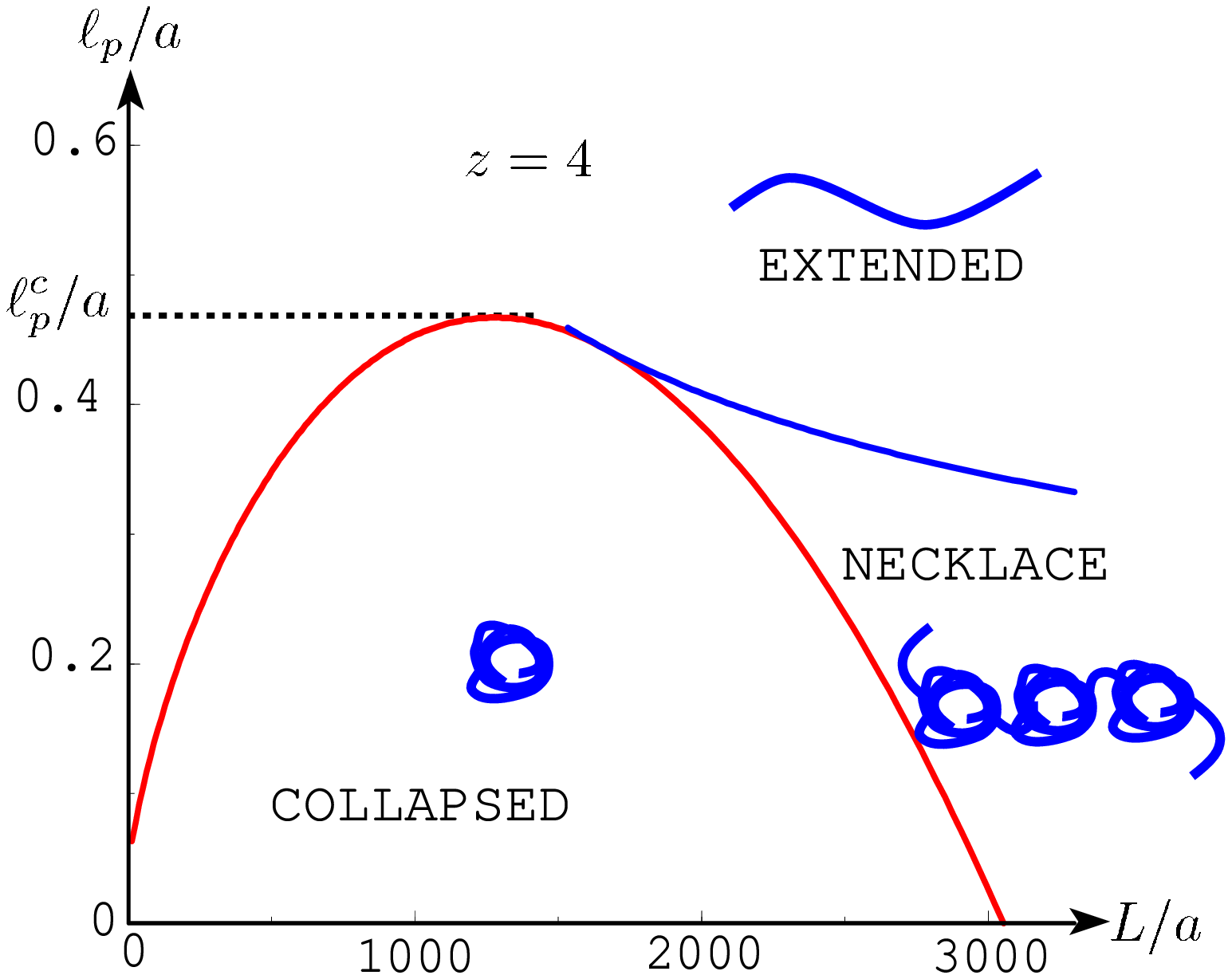}}}\vskip .3truecm
FIG.~2. Phase diagram for the salt-free case, for $z=4$. The parameters 
$a=1.7 \AA$, and $k_0=10^{-3} \pi/L$ are used.
\label{pdsf}\end{figure}

For a low concentration of added salt ($\kappa\ne 0$), the energy cost 
of  deformations is again given by Eq.(\ref{Fns}), but now the function  
${\cal E}(k)$ is analytic, with a well defined power expansion in $k$. 
The first term in the (properly regularized) expansion is $L_pk^4$, 
from which we can define an effective persistence length given by
\begin{equation}
L_p=\ell_p+\frac{q_0^2 \ell_B}
{4\left[1+2(\delta q)^2 (\ell_B/a) \ln(1/\kappa a)\right]^2 (\kappa a)^2}
-{B_2\over 2 \kappa}.   \label{Lp2}
\end{equation}
Once again, $B_2$ is obtained from an integral which has a very slow 
dependence on $(\delta q)^2(\ell_B/a)$, and which for typical values,
can be approximated by $B_2 \simeq c_2/\ln(1/\kappa a)$
with $c_2\simeq 0.288$. 
This leads to the expression for the persistence length in Eq.(\ref{Lpeff}).
It is interesting to note that a similar result has been predicted for stiff
polyampholytes \cite{HaThi}.

The second term in  Eq.(\ref{Lp2}) reproduces the OSF electrostatic
persistence length in the limit $(\delta q)^2=0$\cite{OSF}, with a reduced
charge density $q_0$. 
Upon including the counterion fluctuations, $(\delta q)^2 \neq 0$, there is 
a further reduction in this term.
The final term is a negative contribution coming from the 
fluctuation-induced attractions.
The latter  reduces the  `rigidity' (effective persistence length) of the PE,
which if negative results in  conformational instability (collapse). 
Using this criterion, we obtain the phase diagram of Fig.~3.
As in the salt--free case, the phase boundary has a nearly parabolic shape,
with a maximum at the critical persistence length given in Eq.(\ref{lc1}), 
with $\Delta_{as}=c_2^2$. 
Note that the maximal persistence length is much larger in the presence 
of salt, making it much easier to achieve conditions favorable to collapse.
\begin{figure}
\centerline{\epsfxsize 8cm{\epsffile{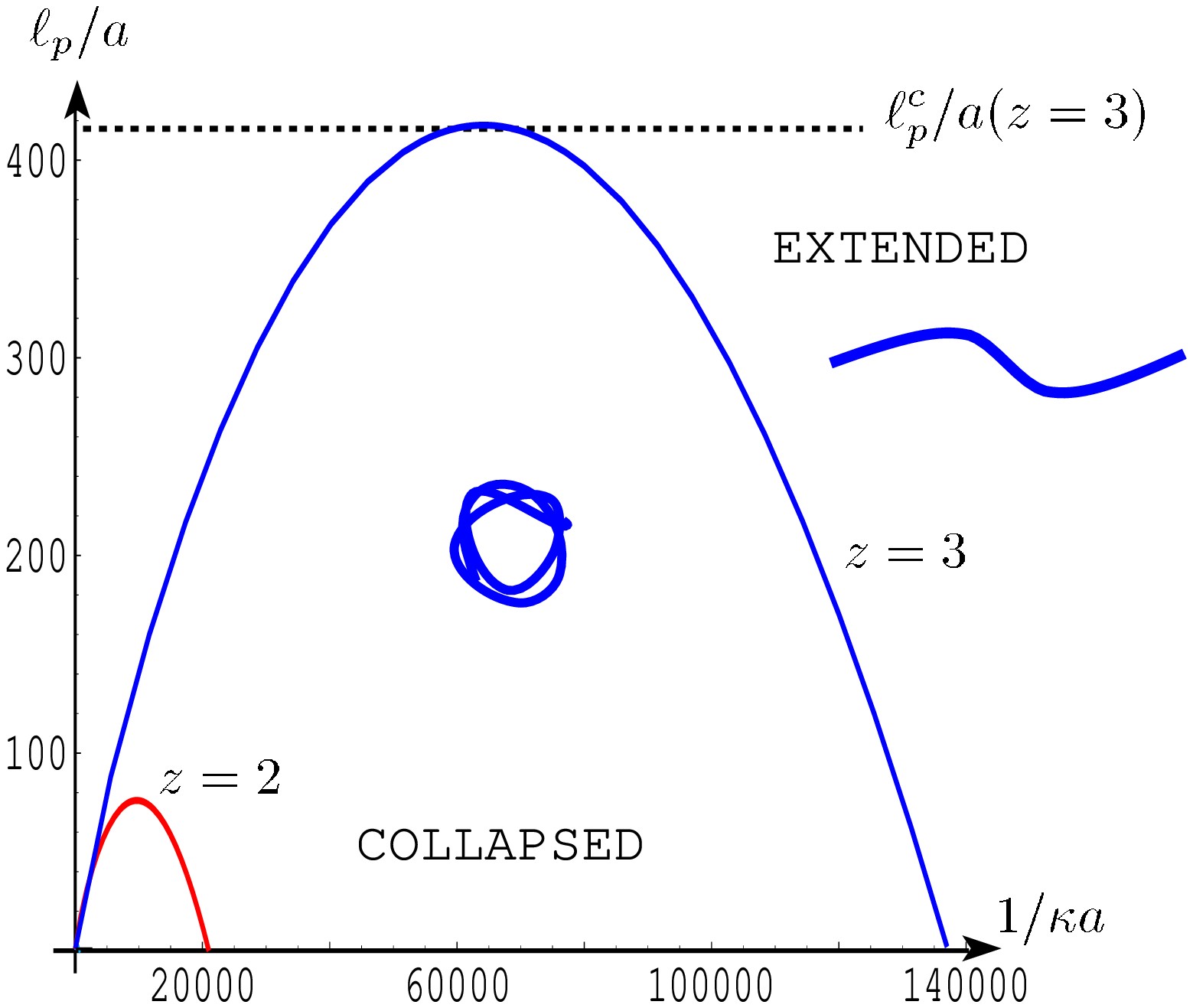}}}\vskip0.3truecm
FIG.~3. Phase diagram for the added-salt case, for different values
of the counterion valence $z$.
\label{pdsa}\end{figure}

Rather than describing the formal steps of the method and approximations
(which will be detailed elsewhere\cite{long}), 
we conclude with a qualitative summary of the nature of the results, 
and the range of their validity.
 From dimensional analysis, it is easy to show that {\it unscreened} Coulomb 
interactions make a contribution of $\ell_B (k/a)^2$ to the
rigidity spectrum ${\cal E}(k)$.
In the PB solution, due to charge condensation the strength of this term is
reduced by a factor of $q_0^2=(a/z\ell_B)^2$.
If the charge density on the PE is allowed to fluctuate, it is further reduced,
and the Coulomb rigidity goes down by a factor of 
$(\delta q)^4(\ell_B/a)^2\ln^2(L/a)$, with $(\delta q)^2=z(1-q_0)$.
However, these reductions do not change the overall sign which still prefers
a rodlike structure.
An attractive (destabilizing) contribution is generated by fluctuation-induced 
interactions, which are typically independent of microscopic parameters,
and hence make a contribution of $-k^3$ to   ${\cal E}(k)$.
Comparing to the leading Coulomb contribution, the latter corrections become
important at {\it short scales} of order of $a^2/\ell_B$.
We thus may well question the applicability of continuum formulations
to describe such a short-distance instability.
In hindsight, the phase diagrams of Fig.~2 and Fig.~3 indicate that the
prefactors involved in softening of the residual repulsion
conspire to make the actual instability lengths quite large 
($\sim z^4 \ell_B^3/a^2$), and thus the continuum formulation should hold for 
a large portion of these phase diagrams \cite{z=1}.
Another potential concern is the relevance of higher order loop corrections
when the instability sets in.
We are not able to address this issue quantitatively, and it may be best to
regard Eq.(\ref{Z2}) as a variational approximation to the full problem.
Finally, the instability analysis performed here only provides us with
information concerning the onset of a conformational change. 
The final structure of the collapsed chain is naturally beyond this
linear stability analysis\cite{Bloomfield,Gelbart}.

It is known experimentally that different counterions with the same valence
may behave differently as collapsing agents. The difference is usually
attributed to modifications in the microscopic structure of the PE
that take place upon binding of the counterions\cite{Bloomfield}. 
It is thus plausible that the microscopic features that distinguish between 
different counterions with the same electrostatic properties can be encoded 
in a single parameter, the microscopic persistence length, which normally 
depends only on the local microscopic structure of the PE backbone.

We have benefited from many helpful discussions with
M. E. Fisher, A. Yu. Grosberg, K. Kremer, and C. R. Safinya.
This research was supported in part by the National Science
Foundation under Grant No. PHY94-07194.
The work at MIT is supported by the NSF grant DMR-93-03667. 
Financial support from the Max-Planck-Gesellschaft (RG and TBL),
and the Iranian ministry of higher education (RG), is also
gratefully acknowledged.

\end{multicols}

\begin{references}


\bibitem[*]{Add1}
Present address

\bibitem{Man}
G.S. Manning, J. Chem. Phys. {\bf 51}, 954 (1969).

\bibitem{Oos}
F. Oosawa, Biopolymers {\bf 6}, 134 (1968); F. Oosawa, {\it Polyelectrolytes}
(Marcel Dekker, New York, 1971).

\bibitem{Delsanti}
M. Delsanti, J.P. Dalbiez, O. Spalla, L. Belloni, and M. Drifford, ACS Symp. 
Ser. {\bf 548}, 381 (1994); 

\bibitem{Monica}
P. Gonzalez-Mozuelos and M. Olvera de la Cruz, J. Chem. Phys. {\bf 103}, 3145
(1995); M. Olvera de la Cruz, L. Belloni, M. Delsanti, 
J.P. Dalbiez, O. Spalla, M. Drifford, J. Chem. Phys. {\bf 103}, 5781 (1995).

\bibitem{Bloomfield}
V.A. Bloomfield, Biopolymers {\bf 31}, 1471 (1991); V.A. Bloomfield, Curr. Opin.
Struct. Biol. {\bf 6}, 334 (1996).

\bibitem{Bundle}
J.X. Tang, S. Wong, P. Tran, and P.A. Janmey, Ber. Bunsen-Ges. Phys. Chem. {\bf 
100}, 1
(1996); J.X. Tang, T. Ito, T. Tao, P. Traub, and P.A. Janmey, Biochemistry {\bf 
36}, 12600
(1997).

\bibitem{Rouzina2}
I. Rouzina and V.A. Bloomfield, J. Phys. Chem. {\bf 100}, 9977 (1996).

\bibitem{Levin}
J.J. Arenzon, J.F. Stlick, and Y. Levin, preprint (1998)(cond-mat/9806358).

\bibitem{Leikin}
A.A. Kornyshev, and S. Leikin, J. Chem. Phys. {\bf 107}, 3656 (1997);
Biophys. J. {\bf 75}, 2513 (1998).


\bibitem{RodFFI}
N. Gronbech-Jensen, R.J. Mashl, R.F. Bruinsma, and W.M. Gelbart, Phys. Rev. 
Lett. {\bf 78}, 2477 (1997); J.L. Barrat and J.F. Joanny, Adv. Chem. Phys. 
{\bf 94}, 1 (1996); R. Podgornik and V.A. Parsegian, Phys. Rev. Lett. 
{\bf 80}, 1560 (1998).

\bibitem{HaLiu}
B.-Y. Ha and A.J. Liu, Phys. Rev. Lett. {\bf 79}, 1289 (1997); B.-Y. Ha and A.J. 
Liu, 
Phys. Rev. Lett. {\bf 81}, 1011 (1998). 

\bibitem{KGRMP}
M. Kardar and R. Golestanian, Rev. Mod. Phys. {\bf 71}, in press (1999)
(cond-mat/9711071).

\bibitem{SK}
M.J. Stevens and K. Kremer, Phys. Rev. Lett. {\bf 71}, 2228 (1993); J. Chem. 
Phys. {\bf 103}, 1669 (1995).

\bibitem{BKK}
N.V. Brilliantov, D.V. Kuznetsov, and R. Klein, Phys. Rev. Lett. {\bf 81}, 1433 
(1998); H. Schiessel and P. Pincus, preprint (1998). 

\bibitem{Man2}
G.S. Manning, Biopolymers {\bf 19}, 37 (1980); G.S. Manning, Cell Biophys. 
{\bf 7}, 57 (1985).

\bibitem{Rouzina}
I. Rouzina, and V.A. Bloomfield, Biophys. J. {\bf 74}, 3152 (1998).

\bibitem{Podgornik}
P.L. Hansen, R. Podgornik, D. Svensek, and V.A. Parsegian, preprint (1998).


\bibitem{LiK}
H. Li and M. Kardar, Phys. Rev. Lett. {\bf 67}, 3275 (1991); H. Li and M. 
Kardar, 
Phys. 
Rev. A {\bf 46}, 6490 (1992). 

\bibitem{OSF}
T. Odijk,, J. Polym. Sci. {\bf 15}, 477 (1977);
J. Skolnick and M. Fixman, Macromolecules {\bf 10}, 944 (1977).

\bibitem{LP}
A.W.C. Lau and P. Pincus, Phys. Rev. Lett. {\bf 81}, 1338 (1998).

\bibitem{RG}
R. Golestanian, Europhys. Lett. {\bf 36}, 557 (1996).

\bibitem{long}
R. Golestanian, M. Kardar, and T.B. Liverpool, in preparation.

\bibitem{fisher}
The density of the salt, $n$, should be low enough not to smear out the 
condensation.
The requirement that the `mass contrast, be substantial, yields 
$n \ll z (1-a/z \ell_B)/(a S)$, where $S$ is the cross sectional area of the PE.

 
\bibitem{k0}
In Ref.\cite{KK}, it is shown that rotational symmetry of the original
Hamiltonian for a charged manifold requires ${\cal E}(k)/k^2 \to 0$, 
in the limit $k \to 0$. For a PE, a cutoff $k_0$ is present as
${\cal E}(k)\sim k^2 \ln(k/k_0)$, and one should take the limit as 
$k \to k_0$ to ensure this requirement. Although, strictly speaking, one 
should finally take the limit $k_0 \to 0$, it is sufficient for 
practical purposes that $k_0$ is small enough such that 
$\sin(k_0 L) \simeq k_0 L$.  
With this in mind, we have selected  $k_0=10^{-3} \pi/L$
for the phase diagram of Fig.~2.



\bibitem{KK}
Y. Kantor and M. Kardar, Europhys. Lett. {\bf 9}, 53 (1989).

\bibitem{LiWi} 
H. Li and T.A. Witten, Macromolecules {\bf 28}, 5921 (1995).

\bibitem{k^3}
The repulsive part of the spectrum comes from a pair potential
of the form $(a/z \ell_B)^2/z^2 \ell_B r$, while the attractive fluctuation-induced part
can be obtained from a $-1/r^2$ pair potential. This leads to a short
length scale instability for $r< z^4 (\ell_B/a)^2 \ell_B$.

\bibitem{necklace}
A necklace structure for randomly charged polymers was introduced in
Y. Kantor and M. Kardar,  Europhys. Lett. {\bf 27}, 643 (1994); 
Y. Kantor and M. Kardar, Phys. Rev. {\bf E51}, 1299 (1995).
Its applicability to uniformly charged PEs is discussed in
A.V. Dobrynin, S.P. Obukhov, and M. Rubinstein, Macromelecules
{\bf 29}, 2974 (1996).

\bibitem{HaThi}
B.-Y. Ha and D. Thirumalai, J. Phys. II (France), {\bf 7}, 887 (1997).

\bibitem{z=1}
 From these arguments, it is clear that counterions with higher 
valence $z$ are much more effective in collapsing the PEs.
Nonetheless, Eq.(\ref{lc1}) does not rule out the possibility of collapse 
for monovalent counterions, provided that the PE has a low enough 
microscopic persistence length. 

\bibitem{Gelbart}
S. Y. Park, D. Harries, and W. M. Gelbart, Biophys. J. {\bf 75}, 714 (1998).



\end{references}
\end{document}